\documentclass{article}
\usepackage{authblk}
\usepackage{amsmath,amsfonts,amssymb} % ams fonts and symbols
\usepackage{graphicx, hyperref} % Required for inserting images
\usepackage[round]{natbib}
\usepackage{comment}
\usepackage{tabularx}
\usepackage{booktabs}
\graphicspath{{./Figs/}} % folder for figures

\newcommand{\mB}{\mathcal{B}}

\newcommand{\mS}{\mathcal{S}}
\newcommand{\mV}{\mathcal{V}}
\usepackage{color}						% supports colored text for editing

\usepackage[normalem]{ulem}			% supports text strikethrough for editing
\DeclareMathOperator*{\argmin}{arg\,min} % thin space between arg and min, limits placed underneath
\date{}
   \title{Implied Probabilities and Volatility in Credit Risk: A Merton-Based Approach with Binomial Trees}
\author[1 *]{Jagdish Gnawali}
\author[2]{Abootaleb Shirvani}
\author[1]{Svetlozar T. Rachev}
\affil[1]{\small Department of Mathematics \& Statistics, Texas Tech University, Lubbock, TX 79409-1042, USA}
\affil[2]{\small Department of Mathematical Sciences, Kean University, Union, NJ 07083, USA}
\affil[*]{Corresponding author:  jgnawali@ttu.edu}

\begin{document}

\maketitle

 \noindent {\bf Abstract}
 
We explore credit risk pricing by modeling equity as a call option and debt as the difference between the firm’s asset value and a put option, following the structural framework of the Merton model. Our approach proceeds in two stages: first, we calibrate the asset volatility using the Black-Scholes-Merton (BSM) formula; second, we recover implied mean return and probability surfaces under the physical measure. To achieve this, we construct a recombining binomial tree under the real-world (natural) measure, assuming a fixed initial asset value. The volatility input is taken from a specific region of the implied volatility surface—based on moneyness and maturity—which then informs the calibration of drift and probability. A novel mapping is established between risk-neutral and physical parameters, enabling construction of implied surfaces that reflect the market’s credit expectations and offer practical tools for stress testing and credit risk analysis.


 \section{Introduction}
The Merton model (\cite{Merton_1974}) simplifies a firm’s capital structure by assuming that the total value of its assets follows a geometric Brownian motion under the physical measure, characterized by a constant mean rate of return and volatility. It assumes no bankruptcy costs, meaning the liquidation value equals the total firm value. Debt and equity are assumed to be frictionlessly tradable. The model considers a firm financed by equity and a single zero-coupon bond with face value $K$ and maturity $T$. At maturity, if the firm’s asset value $V_T$ exceeds $K$, the debt is repaid in full and residual value is allocated to shareholders. If $V_T < K$, default occurs, and bondholders receive the full firm value, while shareholders receive nothing due to limited liability.

While the original model is based on a single maturity, it has often been interpreted to approximate the pricing of other bonds with different maturities and small face values. Merton’s approach uses the Black--Scholes--Merton (BSM) framework to price equity as a European call option on the firm’s assets. Inputs include the current value and volatility of the firm’s assets, the total debt, and the debt’s maturity.

A common implementation, following \cite{Jones_1984}, estimates asset value and volatility from the observed market value and volatility of equity, consolidating all debt obligations into a single maturity payment. \cite{White_2005} proposed an alternative implementation using option-implied volatilities on equity to infer asset dynamics. However, this method depends heavily on option-implied volatilities, which may incorporate market inefficiencies or noise.

In this paper, we propose a novel implementation of the Merton model based on observed asset values, debt maturity, and the prevailing risk-free rate. By treating the current equity price as the value of a European call option on the firm’s assets, we invert the BSM formula to recover the implied asset volatility. This reverse-engineering approach is consistent with the continuous-time Merton framework and offers a more flexible and data-driven calibration method.

To calibrate equity volatility, we use observed equity prices and compare them against market call option prices. Theoretical option prices are computed using the BSM formula, and the implied volatility is obtained by minimizing the squared relative error between observed and model-predicted option prices. We then compare the implied volatilities of assets and equity to examine the relationship between the two.

\cite{Gemmill_2019} assessed the performance of Merton’s model in explaining credit default swap (CDS) spreads, particularly when allowing for time-varying skewness in the risk-neutral distribution of asset returns. Their findings demonstrate that incorporating skewness significantly improves pricing accuracy, especially before and after financial crises. \cite{Jacquier_2015} introduced an expansion of implied volatility using characteristic functions, enabling a refined understanding of implied volatility surfaces across a wide range of financial models, including those with jumps. \cite{Wang_2009} provided a survey of structural credit models beginning with Merton’s, emphasizing extensions that allow for early default via a threshold value approach. In this framework, default occurs when asset value drops below a predetermined barrier, similar to pricing barrier options. This is often interpreted as the result of shareholders’ strategic default behavior. \cite{Lando_2009} and \cite{Loeffler_2011} explored practical applications of structural credit models based on geometric Brownian motion. \cite{Cherubini_2001} incorporated fuzzy measure theory into Merton’s structural credit risk framework to account for liquidity risk, thereby developing a generalized model based on fuzzy set concepts. Numerous contributions by \cite{Black_1976}, \cite{Vasicek_1977}, \cite{Geske_1977}, \cite{Longstaff_1995}, \cite{Klein_1996}, \cite{Lopez_2000}, \cite{Jarrow_2001}, \cite{Duffie_2012}, and \cite{Hull_2009} also develop credit risk models assuming firm value follows a geometric Brownian motion.

Another contribution of this paper lies in estimating real-world asset return and the probability of downward movements, extending beyond the Merton model by using a binomial tree framework. This discrete-time structure enhances realism in the implied volatility surface and remains compatible with the BSM model. \cite{kim_2016} proposed a flexible binomial tree framework that encompasses popular models (CRR, Jarrow–Rudd, Tian) as special cases and fits all moments of the approximated geometric Brownian motion. A special case of their model aligns all tree moments with those of the underlying GBM.

In our setup, the mapping between real-world and risk-neutral parameters becomes explicit. Although this mapping fades in continuous time as $\Delta t \to 0$ (except for volatility), the binomial framework allows us to preserve stylized features—such as drift, skewness, kurtosis, and asymmetric probabilities—under the real-world measure.

A key contribution of this work lies in the calibration of real-world parameters. Building on the one-to-one mapping between physical and risk-neutral probabilities established by \cite{Hu_2020a, Hu_2022}, this relationship was further extended to the trinomial tree framework by \cite{Gnawali_2025}. Based on a European call option price, one can recover the upward movement probability, mean asset return, and volatility. Using the asset volatility recovered from the Merton model, we apply the binomial tree to extract asset return and probability surfaces that match key features of real-world asset dynamics, including asymmetry, drift, and credit stress indicators.

The remainder of the paper is organized as follows. Section  ~\ref{sec: Credit Risk} introduces the credit pricing. Section ~\ref{sec: vol} presents the volatility calibration. Sections~\ref{sec:IV_asset} and \ref{sec:IV_Equity} detail asset and equity volatility calibrations, respectively. Section~\ref{Sec: Binomial} presents the binomial tree model and the link between real-world and risk-neutral parameters. Section~\ref{sec:Im_m, Im_p} and section ~\ref{sec:Eco} present the implied drift and implied probability results. An out-of-sample illustration is presented in section~\ref{sec: OT_sample}.  Section~\ref{sec: conclusion} concludes the paper.

\section{Credit Pricing} \label{sec: Credit Risk}
Suppose an asset $\mV$ consist of two components; a risky bond $\mB $ and a stock $\mS$. At time $t=0$, suppose that the firm has issued debt $B$ (a risky zero-coupon bond) with face value $K$ and maturity date $T\leq\overline{T}$ as well as an equity $\mS$. By taking $K$ as a strike price, the debt can be viewed as the difference between a bond  (which is equivalent to a certain payment of $K$ in the future) and a put option whereas the equity can be viewed as a call option on the assets.  The call option on the asset $\mV$ is the stock price $\mS$ and then payoffs to debt  $B_{T}$ and equity $\mS_{T}$ at date $T$ are given by   
\begin{subequations} \label{eq: equity &  bond}
    \begin{align}
      B{_T}=min(K, V_T) = K-max(K- V_T,0),\\
       \mS{_T} = max(V_T-K,0)
    \end{align}
\end{subequations}
where $V_T$ is the price of asset $\mV$ at time $T$.

Assume that at the terminal time no other parties are receiving any payments from asset $\mV$ and there is no corporate taxes, so that the asset is the sum of debt and equity
\begin{equation}  \label{eq: A=B+S}
   V_T = B{_T} + \mS{_T}
\end{equation}
where $V_T$ is independent of the choice of $K$ \footnote{Modigliani-Miller Theorem: A company's capital structure is not a factor in it's value, market value is determined by the present value of future earnings.}. 

The primary issue at hand is determining the valuation of debt and equity before the maturity date 
$T$. Specifically, how can we accurately price these financial instruments while considering their respective payoff structures and the time remaining until maturity? As we discussed earlier, the equity issued by a firm is equal to a call option, that leads at time $t$,  
the debt is  
\begin{equation}  \label{eq: debt}
B{_t}= \mathcal{D} K -P_t, \quad \mathcal{D}=\frac{1}{R_t}
\end{equation}
where $\mathcal{D}=\frac{1}{R_t}$ is a discount factor and the put option $P_t$ is defined with the help of put call parity 
 $C_t-P_t= V_t- \mathcal{D} K.$
 Therefore the debt is 
 \begin{equation} \label{eq: debt=V_t-C_t}
  B{_t}=(V_t-C_t+P_t)-P_t= V_t-C_t = V_t - S_t.   
 \end{equation}
From  Eq.\eqref{eq: debt=V_t-C_t}, pricing the debt $B_t$ at time $t$ is the difference between asset price $V_t$ and its call option price $S_t$.

\section{Volatility calibration} \label{sec: vol}  
\cite{Merton_1974} is the model that assumes the firm's asset value $\mV$ follows a geometric Brownian motion 
\begin{equation} \label{eq: price dynamics of V}
 dV_t = \mu V_t dt + \sigma V_t d W_t,   
\end{equation}
where $ \mu$ is an asset return, $\sigma$ is an asset volatility and $W_t$ is a standard Brownian motion of asset. Consider $V_0$ be the asset value at $t=0$ then
\begin{equation}
    V_t = V_0 \exp{((\mu-\frac{1}{2}\sigma^2)t+ \sigma W_t).}
\end{equation}

To make the approach more realistic, we consider the asset value $V_t \geq 0$ to represent the value of a certain asset in an economy at time $t$. The Black--Scholes price of a European call option at time $t$ is then given by $C_t(V_t, K, \sigma, r, T - t)$, which is a function of the asset value $V_t$, the strike price $K$ (interpreted as the face value of the debt), the volatility $\sigma$ of the asset price, the risk-free rate $r$, and the time to maturity $T$. According to the Merton model

\begin{equation}  \label{eq: Caliberating sigma}
    S_t = C_t(V_t, K,\sigma, r, T-t).
\end{equation}
 The right-hand side of Eq.\eqref{eq: Caliberating sigma} represents the value of a call option in continuous time under the BSM framework, and thus  yields
 \begin{equation}  \label{eq: Caliberating sigma_2}
      S_t = V_t N (d_1) - K \exp{(-rT(T-t))} N(d_2)
 \end{equation}
 where N is the standard normal distribution function and 
 \begin{subequations}
     \begin{align}
         d_1 = \frac{log(V_t/K)+r(T-t)+\frac{1}{2}\sigma^2 (T-t)}{\sigma \sqrt{T-t}},\\
         d_2 = d_1 - \sigma \sqrt{T-t}.
     \end{align}
 \end{subequations}

We calibrate the volatilities for both the asset and the equity. In Section ~\ref{sec:IV_asset}, we calibrate the asset volatility, and in Section ~\ref{sec:IV_Equity}, we calibrate the equity volatility. To compare the difference between asset and equity volatilities, we fix the maturity time and the moneyness axes for both.

\subsection{ Asset Volatility Calibration} \label{sec:IV_asset}

Asset volatility calibration under the Merton model is distinctive because equity prices are directly interpreted as call option prices within the Black-Scholes-Merton framework, thus informing the underlying asset's volatility. Specifically, we assume a given asset value $V_0$ and consider the adjusted closing price 
$S_t$  of an "Index" at time $t$ as the market price of this call option. This interpretation leverages the Merton model, which treats the firm's asset as composed of a risky bond and equity, where the equity represents a European call option on the asset itself.

Let $G^{(\text{th})} \left( V_0, T, K; \sigma, r_{f,t} \right)$ denote the theoretical call option price calibrated from the BSM model. Under the Merton framework, $S_t$ is interpreted as a call option, where $S_0$ is the adjusted closing price of the equity at time $t = 0$.

The calibration is carried out in continuous time. Under the BSM model, neither the historical mean return nor any return threshold influences the implied volatility. Therefore, the implied asset volatility is obtained by minimizing the squared relative error between the theoretical and observed option prices:

\begin{equation}  \label{eq:IV_Asset}
\sigma_A^{(\text{imp})} = \argmin_\sigma
	\left(
		\frac{G^{(\text{th})} \left( V_0, T, K; \sigma, r_{f,t} \right) - S_0}
			{S_0}
	\right)^2.
\end{equation}

In this study, we adopt a structural credit modeling perspective by interpreting the S\&P 500 index (ticker: SPX) as the equity value of a representative benchmark firm. Within the Merton framework, equity is modeled as a European call option on the firm's assets. Accordingly, we treat the SPX index level as a call option on an assumed fixed asset value $V_0$ with strike corresponding to a hypothetical debt level. This allows us to invert the Black-Scholes-Merton formula and calibrate the implied asset volatility, which we interpret as a market-implied measure of systemic asset risk.

 \begin{table}[htb] 
		\caption{Adjusted closing prices of  SPX for five days on which the option price were recorded.}
    \vspace{0.2cm}
	\label{tab:params}
	\centering
	\begin{tabular}{l ccc ccc cc  cc cc cc l cc }
	\toprule
	Date   & 2/7/2025	& 2/10/2025  & 2/11/2025  & 2/12/2025 & 2/13/2025\\
	\midrule
	Price  & 6025.99 & 6,066.44 & 6,068.50 &  6,051.97 & 6,115.07
   \\
	\bottomrule
	\end{tabular}
    \label{Table:1}
\end{table}

Table~\ref{Table:1} lists the adjusted closing prices of the SPX for five trading days, spanning from February 7, 2025, to February 13, 2025. These equity prices are interpreted as the call option prices of the underlying asset value ($\mV$) on those respective days.

We pull daily closing prices for the SPX and its listed call-option chain from Yahoo Finance, time-stamped February 13, 2025. We retain maturities out to 350 days and strikes spanning 10\%--150\% of spot. The risk-free proxy is the 3-month constant-maturity U.S. Treasury yield from FRED, converted to a daily equivalent. Prior to calibration, we drop option quotes with bids below \$0.05, winsorize implied volatilities at the 99th percentile, and align option expiries to the nearest trading day. All prices are expressed in USD; volatility is annualized. This cleaned five-day snapshot (February 7--13, 2025) underlies both the asset and equity volatility surfaces reported in Figures~1--3.

The asset volatility ($\sigma_A^{(imp)}$) is thus calibrated within the Merton framework in continuous time. We assume a fixed asset value of $V_0 = 1$ trillion dollars\footnote{This assumption is motivated by the interpretation of equity as a call option in the Merton model. Since we consider SPX adjusted closing as equity (i.e., a call option), the asset value must exceed the equity value.}. Using a historical window \( W = \{ t - L + 1, \dots, t \} \)\footnote{The historical window in our computation spans from 02/13/2020 to 02/13/2025.}, the equity price $S_0$\footnote{$S_0$ is the adjusted closing price of SPX on 02/13/2025.} on the final day of the window is taken as the adjusted closing price of the SPX\footnote{The historical data, option prices and the strike prices were downloaded from yahoo finance.}. The corresponding 3 month (converting daily) U.S. Treasury rate on that day is used as the risk-free rate. From Eq.\eqref{eq:IV_Asset}, we calibrate the volatility for the time interval $(t_0, T]$. 

 \begin{figure}[htbp]
	\centering
	\includegraphics[width=0.60\linewidth]{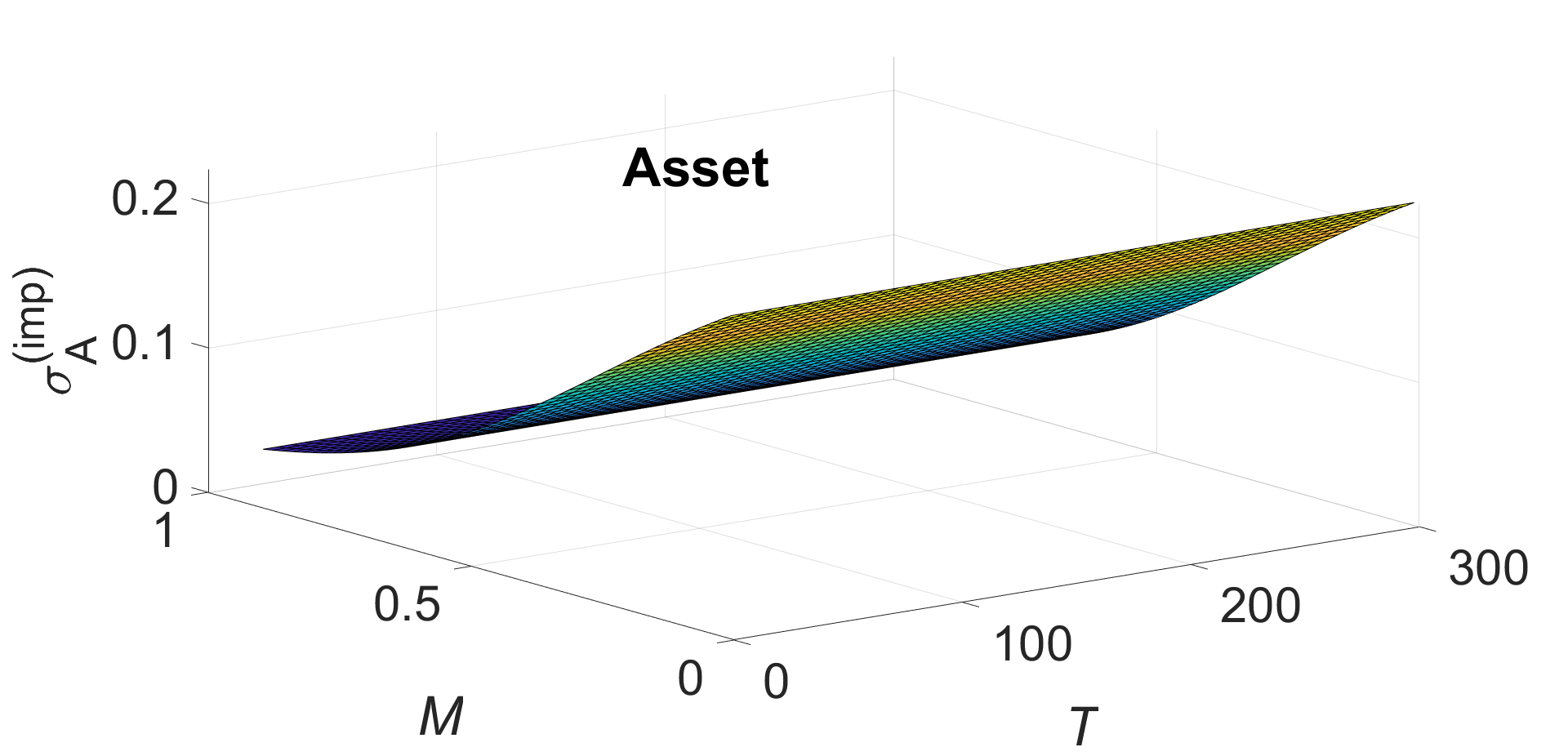}
  \caption{ Asset volatility surface}
  	 \label{fig:Asset_Vol}
\end{figure}

Figure~\ref{fig:Asset_Vol} presents the implied asset volatility surface $\sigma_A^{(\text{imp})}$ as a function of time to maturity $(T)$ (in days) and moneyness $(M = \frac{K}{V_0})$. The surface is calibrated using the Black-Scholes-Merton (BSM) model, covering moneyness from 0.01 (deeply out-of-the-money put options) to $0.9$ (near-the-money options).

The implied volatility surface offers a detailed and nuanced perspective on the market’s assessment of credit risk. In the structural framework where equity is modeled as a European call option on the firm’s assets with strike price equal to the face value of its debt, the implied volatility derived from equity prices captures the uncertainty associated with the firm’s total asset value. When moneyness \( M = \frac{K}{V_0} \) is high—that is, when the debt level \( K \) approaches or exceeds the current asset value \( V_0 \)—the firm is near or in financial distress, and the call option on assets is out-of-the-money. In this region, the implied volatility tends to be lower, reflecting the diminished sensitivity of the option value to volatility changes, despite the elevated credit risk inherent in the firm’s condition. Conversely, when moneyness is low (\( M \ll 1 \)), the firm is well-capitalized, and the equity option is deep in-the-money. Here, implied volatility tends to be higher, primarily due to the option-theoretic properties of deep in-the-money options rather than increased credit risk.

As moneyness approaches unity from below (\( M \rightarrow 1 \)), the implied volatility generally decreases, indicating a transitional regime in which the firm’s solvency is uncertain but not critically impaired. Notably, the observation that the implied asset volatility \(\sigma_A^{(\mathrm{imp})}\) remains below 0.03 across all maturities near \( M \approx 1 \) suggests a market consensus of relative stability in the firm’s asset value and moderate expectations regarding its variability. This behavior aligns with economic intuition regarding credit risk embedded in option-implied measures.

Therefore, this volatility surface provides insight not only into the statistical dispersion of returns but also into the structural credit risk embedded in the firm’s capital structure. Elevated implied volatility in low-moneyness regimes signals potential financial distress, informing both debt pricing and risk management strategies. In contrast, lower volatility near-the-money reflects stable asset coverage and reinforces investor confidence in the firm’s financial soundness.

\subsection{Equity Volatility Calibration}  
\label{sec:IV_Equity}

The goal of this section is to estimate and visualize the implied equity volatility surface $\sigma_E^{(\text{imp})}$ by calibrating the Black-Scholes-Merton (BSM) model to match observed market prices of SPX call options. Using the collected option price data from Yahoo Finance and assuming a total asset value of $V_0 = 1$ trillion dollars, we minimize the squared relative error between theoretical and empirical option prices to derive the implied volatility across different strike prices and maturities. This process yields a surface that provides a detailed view of how the market prices equity risk, offering a critical benchmark for comparison against the model-derived asset volatility surface and enabling deeper insights into the dynamics of market-implied uncertainty.

The implied equity volatility surface plays a central role in our framework by offering an empirical anchor. While the asset volatility calibration is rooted in the Merton structural model, the equity volatility calibration is directly extracted from market-observed option data, reflecting investor pricing of risk and uncertainty tied to SPX equity options. By analyzing the similarities and differences between the two volatility surfaces, we gain important insights into the relationship between firm-level asset risk and market-perceived equity risk, sharpening our understanding of credit risk pricing, volatility transmission, and the alignment (or misalignment) between theoretical models and real-world market dynamics.

Let us consider the historical window described in Section~\ref{sec:IV_asset}, and let $S_t$ represent the adjusted closing price of the SPX, treated as the equity price. Let $G^{(\text{emp})}(S_0, T, K)$ denote the observed market prices of call options written on the underlying equity $\mathcal{S}$, with maturity $T$ and strike prices $K$. Similarly, let $G^{(\text{th})}(S_0, T, K; \sigma, r_{f,t})$ denote the corresponding theoretical call option prices calculated under the BSM model, where $\sigma$ is the volatility and $r_{f,t}$ is the risk-free rate at time $t$. As noted in Section~\ref{sec:IV_asset}, $S_0$ is the adjusted closing price on the final day of the historical window, taken as $6{,}115.07$, with the interest rate $r_{f,0}$ as the 3-month (converted to daily) U.S. Treasury yield at time $t = 0$. The implied equity volatility is thus obtained by minimizing the squared relative error between theoretical and observed option prices:

\begin{equation}  
\label{eq:IV_Equity}
\sigma_E^{(\text{imp})} = \argmin_\sigma
\left(
\frac{G^{(\text{th})} \left( S_0, T, K; \sigma, r_{f,t} \right) - G^{(\text{emp})}(S_0, T, K) }
{G^{(\text{emp})}(S_0, T, K)}
\right)^2.
\end{equation}

\begin{figure}[htbp]
\centering
\includegraphics[width=0.48\linewidth]{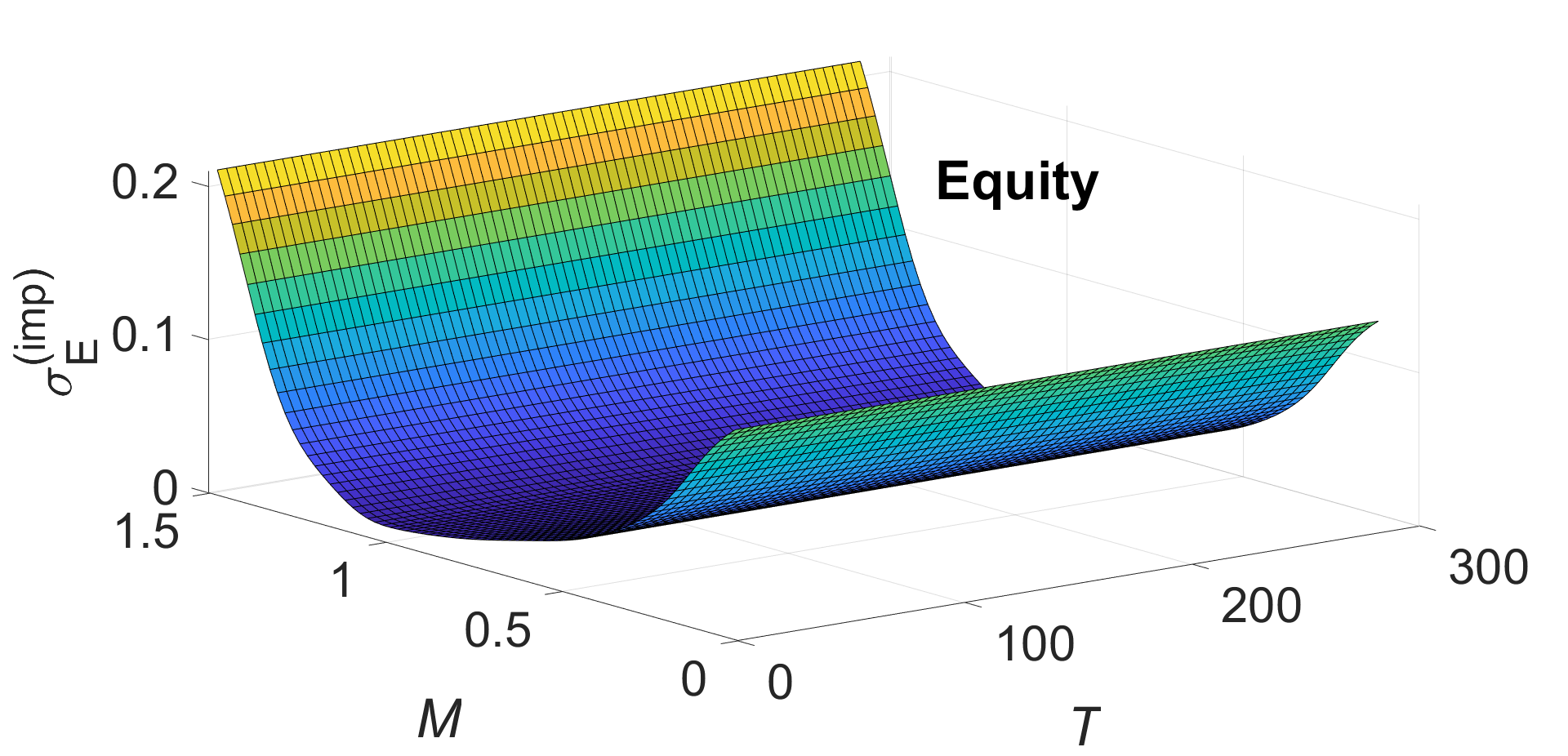}%
\includegraphics[width=0.48\linewidth]{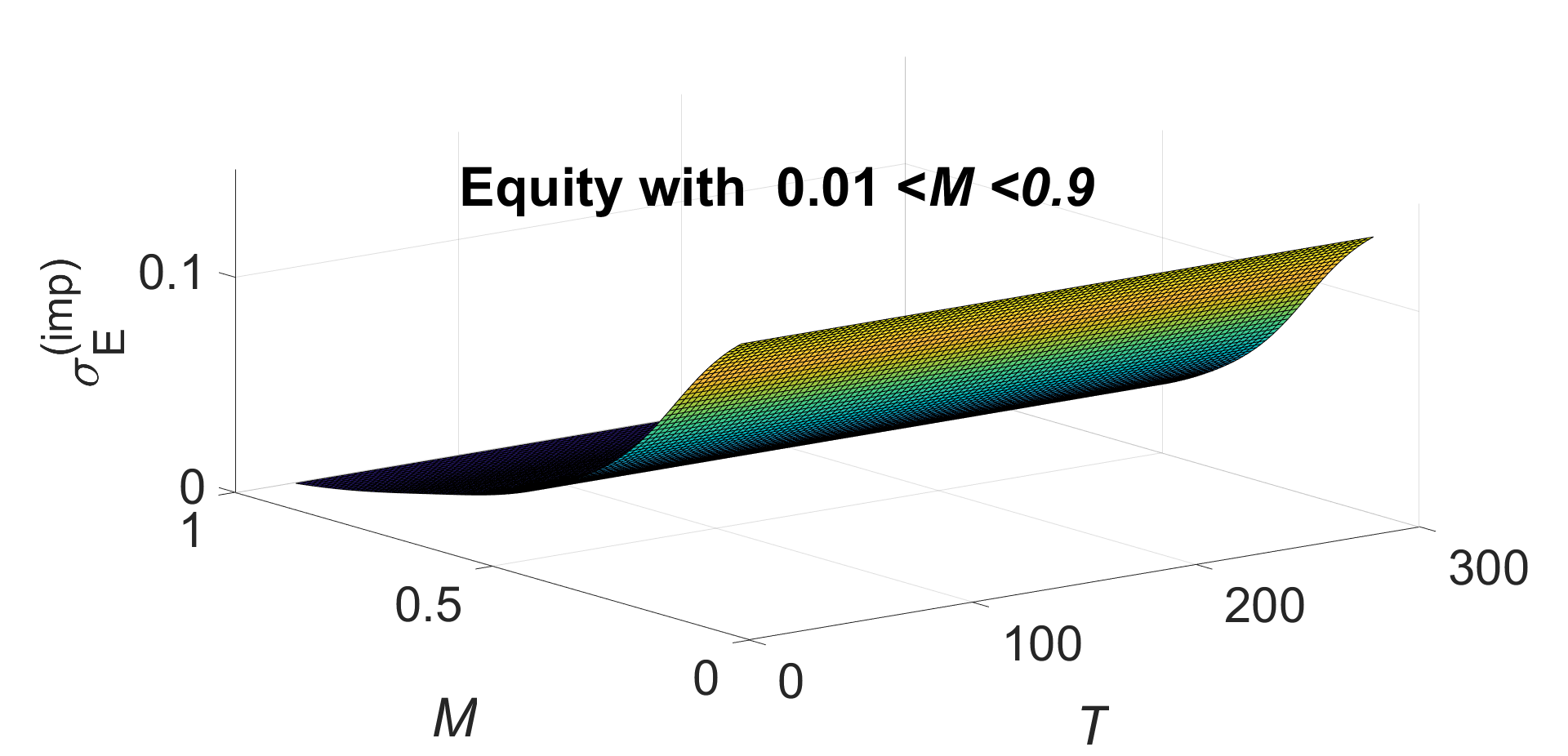}%
\caption{Equity implied volatility surface. For moneyness between 0 and 1.5 (left) and between 0 and 1 (right).}
\label{fig:Equity_Vol}
\end{figure}

Figure~\ref{fig:Equity_Vol} presents the implied equity volatility surface $\sigma_E^{(\text{imp})}$ as a function of time to maturity $T$ and moneyness $M = \frac{K}{S_0}$, where $S_0$ is the spot price and $K$ is the strike price. The left panel shows the full moneyness range $0 < M < 1.5$, while the right panel focuses on $0 < M < 0.9$, aligning with the asset volatility analysis. The data spans five trading days, February 7--13, 2025, covering all unique observed option maturities.

This implied equity volatility surface provides a visual map of how market participants perceive and price equity-related risk. Implied volatility rises as moneyness exceeds 1, reflecting sensitivity to out-of-the-money call options, and declines near the money, where uncertainty is perceived as lower. Comparing this to the asset volatility surface reveals the difference between total firm-level risk and pure equity market risk.

\begin{figure}[htbp]
\includegraphics[width=0.55\linewidth]{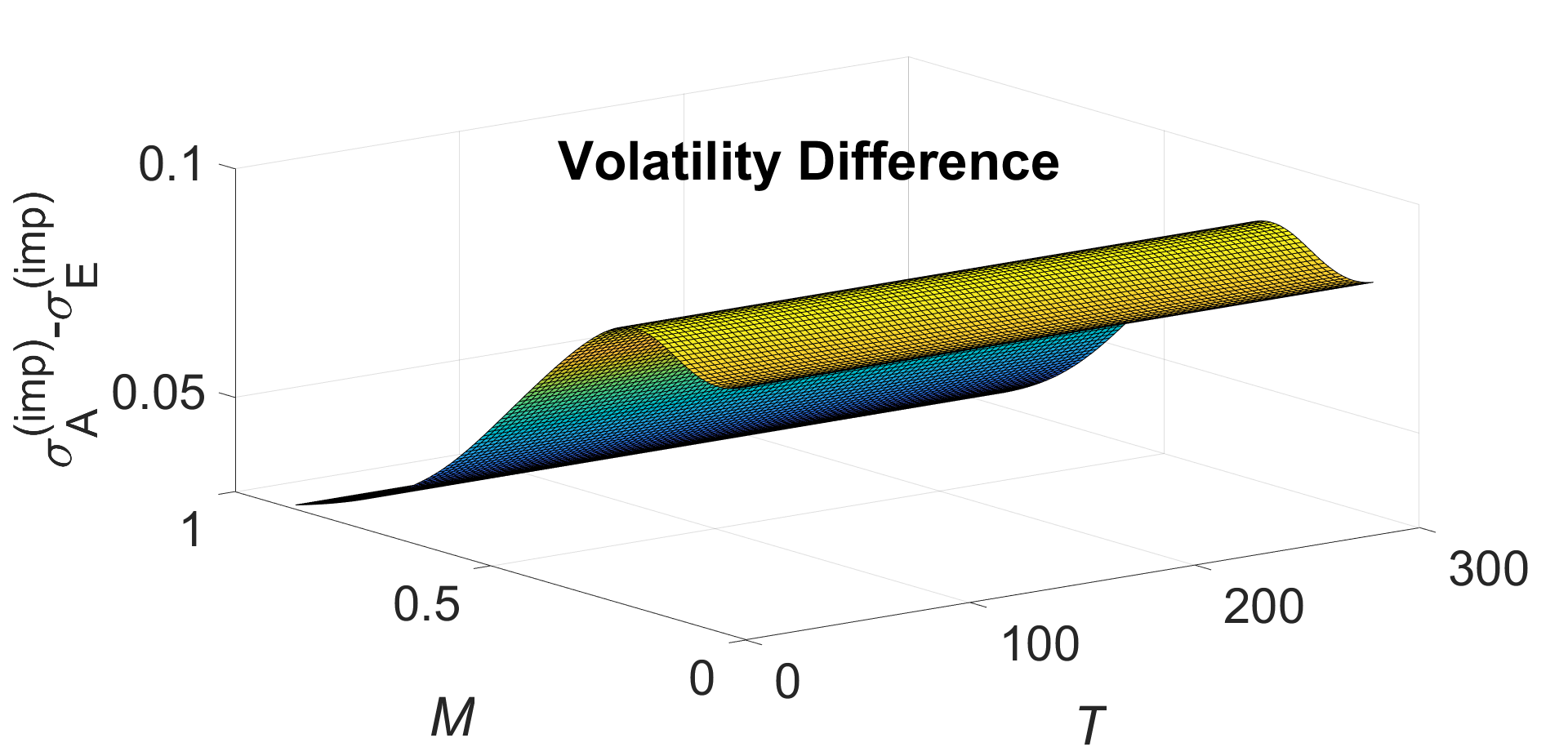}%
\includegraphics[width=0.55\linewidth]{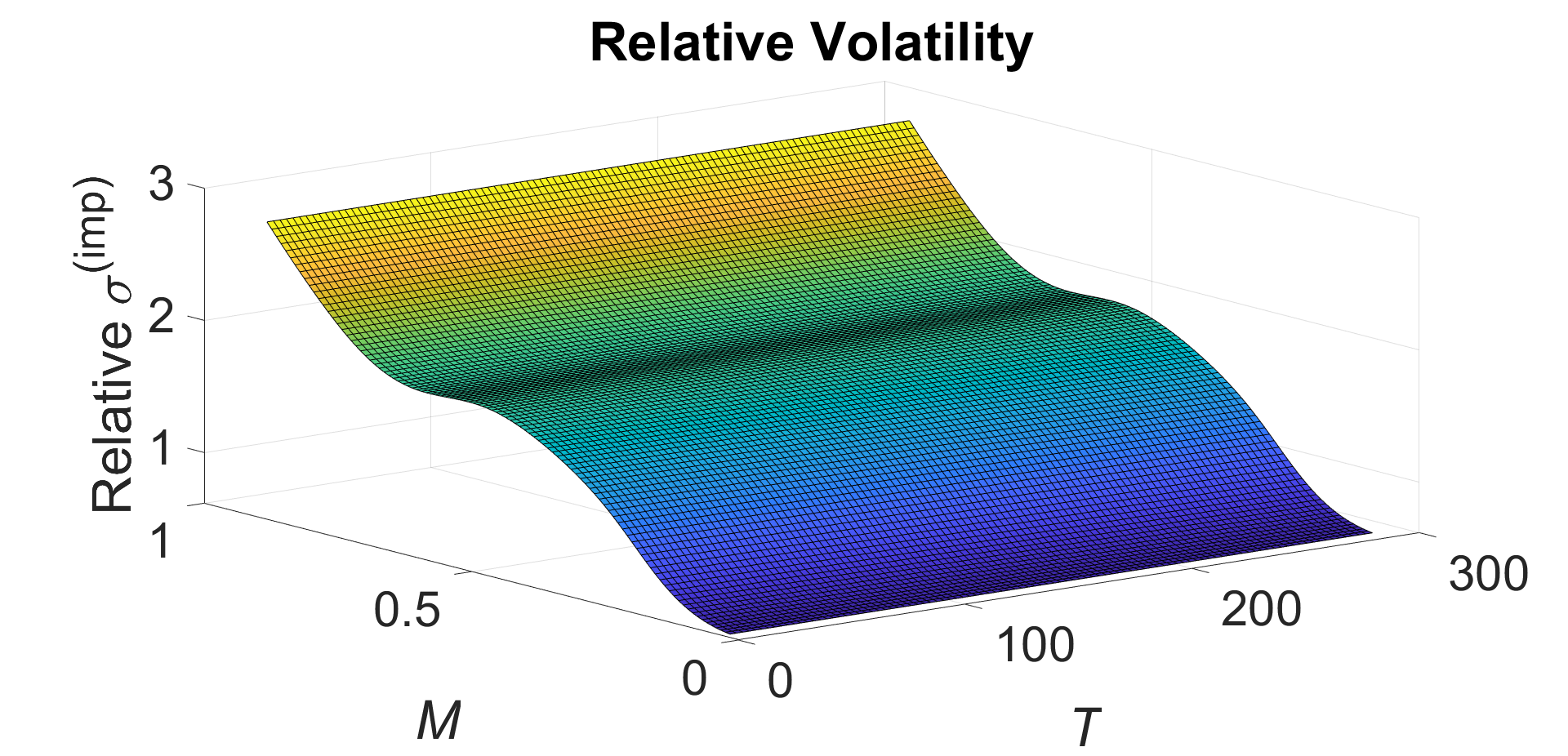}%
\caption{Difference (Asset-Equity) implied volatility surface (left), and relative difference volatility surface (right).}
\label{fig:Diff_Relative_Imp_Vol}
\end{figure}

The left panel of Figure~\ref{fig:Diff_Relative_Imp_Vol} presents the surface of the absolute difference between asset and equity implied volatilities, $\sigma_A^{(\text{imp})} - \sigma_E^{(\text{imp})}$. Across all moneyness levels and maturities, this difference remains strictly positive, with values exceeding 0.029. The minimum difference occurs near $M \rightarrow 1$, indicating that equity and asset volatilities are most aligned when options are near-the-money. However, as moneyness decreases—i.e., as strike prices fall further below the current equity price—a progressive divergence emerges, culminating in a pronounced gap for deep out-of-the-money puts (i.e., as $M \rightarrow 0$). This widening gap reflects the growing influence of credit risk in equity valuation under structural models: while asset volatility captures the total uncertainty of firm value, equity acts as a residual claim and behaves like a call option on the firm's assets. In distressed regimes, where asset values flirt with the debt threshold, equity becomes increasingly sensitive to downside shocks. Yet paradoxically, due to the nonlinear payoff structure, equity’s implied volatility does not rise as sharply as that of the underlying assets. The observed premium in asset volatility—peaking near 0.09 in low-moneyness regions—therefore signals market pricing of heightened default risk, where asset fluctuations pose significant implications for solvency, even as equity retains limited upside exposure.

The right panel of Figure~\ref{fig:Diff_Relative_Imp_Vol} depicts the relative difference, defined as $\frac{\sigma_A^{(\text{imp})} - \sigma_E^{(\text{imp})}}{\sigma_E^{(\text{imp})}}$. This surface captures the proportional volatility premium embedded in the asset view compared to equity. Interestingly, the relative difference declines monotonically as moneyness decreases, with values exceeding 3 near $M = 0.9$ and falling to approximately 0.6163 when $M = 0.032$. This suggests that although the absolute gap remains large in deep out-of-the-money regions, equity implied volatility becomes more reactive in proportional terms. Such behavior highlights the convexity effects in equity valuation: as equity approaches near-worthlessness (i.e., in extreme distress), its sensitivity to marginal changes in asset value increases due to the "optionality" of limited liability. Consequently, while asset volatility remains elevated to reflect firm-wide uncertainty, equity volatility increases more rapidly in regions where option prices are highly sensitive to small changes in the underlying asset, despite a low probability of ending in-the-money. The diminishing relative premium thus reveals a non-linear transmission of risk from the firm's balance sheet to equity pricing, offering insight into capital structure dynamics under stress. These results underscore the importance of distinguishing between asset- and equity-based volatility surfaces in the calibration of credit-sensitive models.

\section{Binomial Model}  \label{Sec: Binomial}

The scenario in the real world is more closely aligned with a discrete structure. For a fixed maturity time $T$, suppose that $V_{k\Delta, n}$, for $k = 0, 1, 2, \dots, n$, where $n \in \mathbb{N}$ and $\Delta = \frac{T}{n}$, represents the asset price at time $k\Delta$. For every $n \in \mathbb{N}$ and for $k = 1, 2, \dots, n$, let $\xi_{k,n}$ be i.i.d. Bernoulli random variables such that the price change probabilities in the natural world satisfy $p(\xi_{k,n} = 1) = 1 - p(\xi_{k,n} = 0)$. The pricing trees in the binomial model are adapted to the discrete filtration
\[
\mathbb{F}^{(n)} = \left\{ \mathcal{F}_k^{(n)} = \sigma(\xi_{j,n} : j = 1, 2, \dots, k),\ k = 1, 2, \dots, n;\ \mathcal{F}_0^{(n)} = \{ \emptyset, \Omega \},\ \xi_{0,n} = 0 \right\}
\]
on a complete probability space $(\Omega, \mathcal{F}, \mathbb{P})$. The asset price dynamics $V$ follows:

\begin{equation}\label{eq:Bino_tree}
	V_{(k+1)\Delta, n} = 
	\begin{cases} 
	     u_{k\Delta,n} V_{k\Delta, n},   & \text{with probability } p_{u,k\Delta},\\
	     d_{k\Delta,n} V_{k\Delta, n},   & \text{with probability } 1 - p_{u,k\Delta}.
	\end{cases}
\end{equation}

For arithmetic returns, the parameters are\footnote{To simplify notation, we write $u$ for $u_{k\Delta,n}$, $d$ for $d_{k\Delta,n}$, and $p$ for $p_{u,k\Delta}$.}
\begin{equation}\label{eq:artn}
	u = 1 + U,\qquad d = 1 + D,  \qquad  R = 1 + r,
\end{equation}
while for log-returns, the parameters are modeled as
\begin{equation}\label{eq:lrtn}
	u = e^{U}, \qquad d = e^{D}, \qquad R = e^{r\Delta}.
\end{equation}
Here, $r$ denotes the risk-free rate. \cite{Hu_2020a} addressed the misconception that binomial option pricing is independent of the return $\mu$ and the probability $p$, and estimated the parameters as
\begin{subequations} \label{eq: U and D}
\begin{align}
 	U &= \mu \Delta + \sigma \sqrt{\frac{1-p}{p}} \sqrt{\Delta},\\
 	D &= \mu \Delta - \sigma \sqrt{\frac{p}{1-p}} \sqrt{\Delta}.
\end{align}
\end{subequations}

This leads to the option price under the no-arbitrage argument:
\begin{equation} \label{eq: option_price}
    f_k = \frac{1}{R} \left( q f_{k+1}^u + (1 - q) f_{k+1}^d \right),
\end{equation}
where
\begin{equation} \label{eq:risk neutral Probability}
  q = p - \theta \sqrt{p(1 - p)\Delta}
\end{equation}
is the risk-neutral probability, with the risk-reward ratio
\begin{equation}
  \theta = \frac{\mu - r}{\sigma}.
\end{equation}

The call option price formula in Eq.~\eqref{eq: option_price} can be further written as
\begin{equation}   \label{eq: option_price_1}
    \begin{aligned}
    S_k &= \frac{1}{R} \left(p - \frac{\mu - r}{\sigma} \sqrt{p(1 - p) \Delta} \right) f_{k+1}^u \\
        &\quad + \frac{1}{R} \left(1 - p + \frac{\mu - r}{\sigma} \sqrt{p(1 - p) \Delta} \right) f_{k+1}^d,
   \end{aligned}
\end{equation}
where the payoffs are given by $f_{k+1}^u = \max(u V_k - K, 0)$ and $f_{k+1}^d = \max(d V_k - K, 0)$.

Using Eq.~\eqref{eq: U and D}, we construct a theoretical asset pricing tree in Eq.\eqref{eq:Bino_tree} with unknown parameters $\mu$ and $p$, while the volatility parameter $\sigma$ is calibrated in Section~\ref{sec: vol}.

\subsection{ Asset Implied Mean and  Implied Probability} \label{sec:Im_m, Im_p}
We define a window \( W = \{ t - L + 1, \dots, t \} \) of historical data, as described in Section~\ref{sec:IV_Equity}. For each \( t \in W \), we use the adjusted daily closing price of the SPX to represent the call option price within that window. We assume the asset value to be \( V_0 = 1 \) trillion dollars, and the risk-free rate is taken as the U.S. Treasury three-month yield (converted to a daily rate) for each day within the window.

For each day \( t \in W \), we exclude any dates where the adjusted closing price is unavailable. We define the following two sets:
\[
R_f = \{ r_t : r_t \text{ is the risk-free rate on day } t \}, 
\]
 \[
AC = \{ AC_t : AC_t \text{ is the adjusted closing price on day } t \}.
\]

For each \( t \in W \), we verify the existence of both the risk-free rate \( r_t \in R_f \) and the adjusted closing price \( AC_t \in AC \). If \( r_t \) is missing for a given day \( t \) but \( AC_t \) is available, we search for the nearest available day within \( W \) where \( r_t \) exists and use it to impute the missing value for day \( t \). After this imputation process, we ensure that the sets \( R_f \) and \( AC \) are of equal length.

Let \( T \) denote the asset maturity grid. The call option prices are computed starting from the date \( t = L - T \), using Eq.~\eqref{eq: option_price}. The mean return surface is then calibrated by minimizing the following objective function:

\begin{equation}\label{eq:IM}
\mu^{(\text{imp})} = \arg\min_\mu
	 \left (
		\frac{ G^{(\text{th})}(V_0,T,K, \sigma, r_{f,t}, p, \mu ) - S_t }
			{S_t}
	\right )^2.
\end{equation}

Similarly, the implied probability surface is obtained by:

\begin{equation}\label{eq:IP}
p^{(\text{imp})} = \arg\min_p
	 \left (
		\frac{ G^{(\text{th})}(V_0,T,K, \sigma, r_{f,t}, p, \mu ) - S_t }
			{S_t}
	\right )^2,
\end{equation}
where \( G^{(\text{th})}(V_0, T, K, \sigma, r_{f,t}, p, \mu) \) denotes the theoretical call option price computed from the binomial tree model.

\subsection{Economic Interpretation of Real-World Drift and Probability Surfaces} \label{sec:Eco}
This subsection focuses on calibrating two real-world parameters—the implied mean return $\mu^{(\text{imp})}$ and the implied probability of an upward move $p^{(\text{imp})}$—to better understand the directional expectations embedded in market option prices. These parameters are estimated using a binomial tree framework, with the previously calibrated asset volatility surface $\sigma_A^{(\text{imp})}$ serving as an input.

To isolate each parameter, we adopt distinct calibration strategies. When estimating the implied mean return, we fix the probability parameter at $p = 0.5$, consistent with a neutral market assumption under the Efficient Market Hypothesis (EMH). This approach allows us to extract the market’s directional drift without imposing a bias in likelihood. Conversely, when calibrating the implied probability, we hold the mean return fixed at $\mu = 0.08$, representing an 8\% annualized growth rate in line with historical asset performance. The volatility used in both calibrations is selected from deep out-of-the-money(for put option) regions of the implied volatility surface (e.g., $M = 0.01$), ensuring that the analysis captures market perceptions of rare but impactful downside scenarios.
\begin{figure}[htbp]
	\centering
	\includegraphics[width=0.48\linewidth]{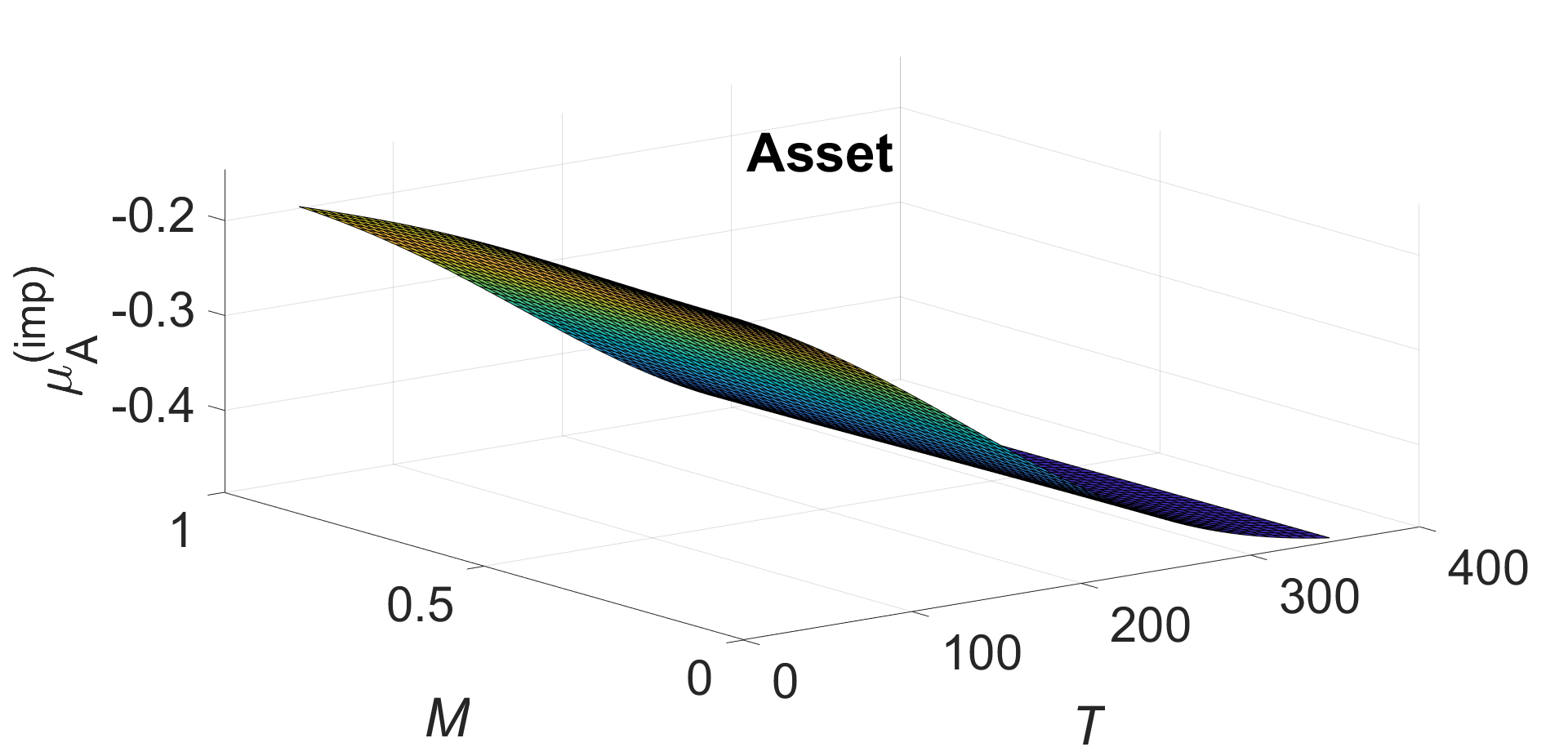}%
	\hfill
	\includegraphics[width=0.48\linewidth]{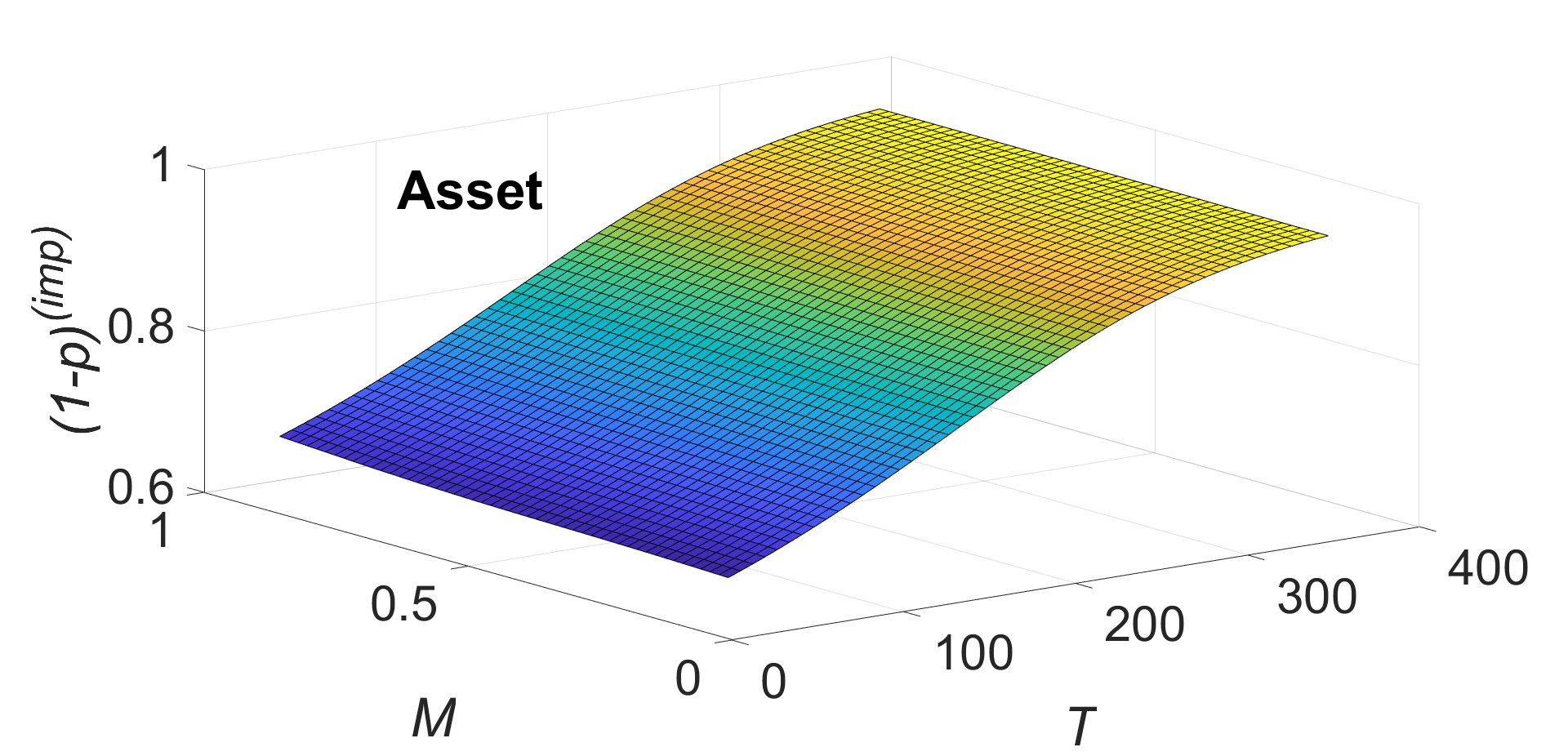}
	\caption{Implied surfaces: asset return (left) and downward probability (right).}
	\label{fig:mu}
\end{figure}

The left panel of Figure~\ref{fig:mu} displays the implied mean return surface under the physical measure. Across all maturities and moneyness levels, $\mu^{(\text{imp})}$ remains persistently negative, including near-the-money and short-dated regions. This is not an artifact of noise or calibration error; rather, it reflects a consistent pricing of adverse expectations in the physical measure. In particular, the mean return $\mu^{(\text{imp})}$ becomes significantly negative in regions where the moneyness $M = \frac{K}{V_0}$ is high—that is, where the strike price far exceeds the asset value, indicating that the firm’s asset value is well below its debt obligation. This reflects the market’s expectation of downward asset drift under financial distress. As moneyness approaches 1 (when the firm is marginally solvent), the drift becomes less negative, and in regions where $M < 0.3$, it levels off, consistent with greater confidence in solvency. The persistent negativity of the inferred drift across the surface signals that the market embeds a substantial premium for credit-related concerns, possibly due to heightened default risk, solvency challenges, or deteriorating firm fundamentals.

The right panel of Figure~\ref{fig:mu} presents the implied probability of a downward move, defined as $1 - p^{(\text{imp})}$. This surface reveals a strong asymmetry in the distribution of expected outcomes. The probability remains above 0.63 even for short maturities and rises sharply with time to maturity, reaching nearly 0.98 as $T \approx 400$. In high moneyness regions (i.e., high $K$), the market assigns over 90\% probability to asset declines. This is consistent with structural credit risk interpretation: as $K$ increases, the firm’s distance to default narrows, and downside risk becomes dominant. The sustained elevation of downside probability across the moneyness–maturity grid reflects the market’s expectations of both immediate and long-term credit deterioration.

Together, the implied mean return and probability surfaces reveal a market that prices the asset under a physical measure dominated by credit concerns. The combination of persistently negative drift and elevated downside probability suggests that investors anticipate not just losses, but credit-driven declines linked to solvency risk, funding instability, and potential default. These distortions in the implied dynamics reflect structural credit stress rather than mere volatility effects, offering a forward-looking signal of deterioration even in the absence of explicit default events.

\subsection{Out-of-Sample Illustration} \label{sec: OT_sample}

\textbf{Stress metric.} We monitor the implied downside probability at moneyness $M = 0.9$ and a 30-day maturity, and compare it with the VIX. Notable spikes in the modeled probability coincide with major stress events, including the COVID-19 market sell-off (March 2020) and the regional banking turmoil (March 2023), lending support to its role as a forward-looking credit-risk indicator.

\textbf{Back-test illustration.} A stylized long/short overlay strategy—reducing equity exposure when the implied downside probability exceeds 80\%, and increasing it when the probability falls below 60\%—shows potential to mitigate downside risk and enhance risk-adjusted performance. While exact improvements in drawdown and Sharpe ratio are not calibrated here, the visual alignment of probability spikes with major market stress events suggests meaningful forward-looking informational content.

\section{Conclusion} \label {sec: conclusion}

This study builds on the structural foundation of the Merton model by interpreting equity as a European call option on the firm’s total asset value. Using observed equity prices and the Black-Scholes-Merton formula, we recovered the implied asset volatility and constructed a recombining binomial tree under the physical (real-world) measure. From this, we derived implied drift and implied probability surfaces that capture the market’s expectations of future asset dynamics.

The results reveal persistently negative drifts and elevated downside probabilities—clear indicators of credit risk embedded in equity prices. These findings suggest that the market anticipates not only adverse asset outcomes, but also credit-driven deterioration, even in the absence of observable default events. A key contribution of this work is the formal mapping between risk-neutral and physical parameters, enabling the recovery of economically meaningful credit risk indicators from derivative-implied information. Applied to the S\&P 500 Index (SPX) with a stylized firm value, the proposed framework offers a practical tool for identifying market-implied credit concerns, which may support stress testing, early warning signals, and credit risk monitoring efforts.

\bibliographystyle{agsm}
\bibliography{Arxiv_06_14.bib}

\end{document}